**COVID-19 Detection and Analysis From Lung CT Images using Novel Channel Boosted CNNs**

Saddam Hussain Khan [1*]

[1]Department of Computer Systems Engineering, University of Engineering and Applied Science, Swat 19060, Pakistan

Corresponding author: (e-mail: saddamhkhan@ueas.edu.pk).

## Abstract

In December 2019, the global pandemic COVID-19 in Wuhan, China, affected human life and the worldwide economy. Therefore, an efficient diagnostic system is required to control its spread. However, the automatic diagnostic system poses challenges with a limited amount of labeled data, minor contrast variation, and high structural similarity between infection and background. In this regard, a new two-phase deep convolutional neural network (CNN) based diagnostic system is proposed to detect minute irregularities and analyze COVID-19 infection. In the first phase, a novel SB-STM-BRNet CNN is developed, incorporating a new channel Squeezed and Boosted (SB) and dilated convolutional-based Split-Transform-Merge (STM) block to detect COVID-19 infected lung CT images. The new STM blocks performed multi-path region-smoothing and boundary operations, which helped to learn minor contrast variation and global COVID-19 specific patterns. Furthermore, the diverse boosted channels are achieved using the SB and Transfer Learning concepts in STM blocks to learn texture variation between COVID-19-specific and healthy images. In the second phase, COVID-19 infected images are provided to the novel COVID-CB-RESeg segmentation CNN to identify and analyze COVID-19 infectious regions. The proposed COVID-CB-RESeg methodically employed region-homogeneity and heterogeneity operations in each encoder-decoder block and boosted-decoder using auxiliary channels to simultaneously learn the low illumination and boundaries of the COVID-19 infected region. The proposed diagnostic system yields good performance in terms of accuracy: 98.21 %, F-score: 98.24%, Dice Similarity: 96.40 %, and IOU: 98.85 % for the COVID-19 infected region. The proposed diagnostic system would reduce the burden and strengthen the radiologist's decision for a fast and accurate COVID-19 diagnosis.

**Keywords:** COVID-19, CT Lung, Detection, Analysis, Boosting, CNN, Split-Transform-Merge, Transfer-Learning, and CNN.

## 1.    Introduction

The coronavirus (COVID-19), named SARS-CoV-2, is a pathogenic viral infection that originated in Wuhan (China) in December 2019 and spread worldwide [1]. COVID-19 causes a respiratory illness and can transit across fever, cough, myalgia, pneumonia, organ failure like the kidney, heart attack, etc. [2]. The total number of COVID-19 cases is approximately 618 million, with 6.6 million deaths, while 597 million have been recovered. COVID-19 is a global pandemic that has devastatingly affected world health and the economy [3]. Therefore, diagnostic tests are required to identify COVID-19 infected patients and reduce the spread timely.

The generally used tests for assessing COVID-19 individual is genetic sequencing and radiological imaging techniques [4]. Moreover, the Real-time Polymerase Chain reaction (RT-PCR) is used to detect SARS-CoV-2. However, RT-PCR testing kits are expensive, and having limited availability and sampling capacity has led to the developing of more efficient techniques [5]. Therefore, the radiological images (X-Ray, Computed Tomography (CT)) analysis has been used as a detection tool to tackle false-negative PCR in suggestive patients. Moreover, the CT scan is utilized for severity assessment, clinical evaluation, monitoring, and treatment of COVID-19 patients [6].

In a public health emergency, the manual examination of many CT images is a great challenge and a severe concern for remote areas without experienced radiologists [7]. There is an intense need for a computer-based tool that can help radiologists in improving performance and deal with many patients [8], [9]. Therefore, deep learning (DL)-based diagnostic systems are developed to facilitate radiologists in identifying COVID-19 infection. Convolutional Neural Network (CNN) is the branch of DL that is considered a powerful tool for medical diagnosis [10]. Moreover, the deep CNN tools can detect minor irregularities that cannot be observed through a manual examination. An effective diagnostic system can overcome the radiologist's burden for manual assessment of COVID-19 images, ultimately improving the survival rate.

The correct analysis of COVID-19 infected images is challenging due to (i) minor contrast variation between the infected and the background boundaries, (ii) high texture variation within homogeneous infected regions, and (iii) the COVID-19 infected region has a high variation in size, shape, and position. Moreover, these radiological images are normally complex in nature and highly distorted due to noise during CT image acquisition [11]. Furthermore, lungs images manifests COVID-19 specific radiological patterns that are usually categorized by various types of opacities: ground-glass-opacities (GGO), reticulation, and pleural [12].

This work addressed the aforementioned challenges and presented a new deep CNN-based framework to detect COVID-19-specific features and analyze thoracic radiologic images. These deep CNN-based systems can capture useful dynamic features of the infected regions, discriminating the COVID-19 infected region from the healthy ones. The significant contributions of the proposed diagnosis system are as follows:

1. A new two-phase deep CNN-based diagnostic system is developed for detecting COVID-19 infection and analyzing suspicious lesions in CT Lungs images to identify the severity and stage of the disease.

2. A novel SB-STM-BRNet CNN comprised of a new dilated convolutional Split-Transform-Merge (STM) block, and channel Squeezed and Boosted (SB) idea is developed to screen COVID-19 infected images. The STM blocks performed multi-path region-smoothing and boundary operations to learn minor contrast variation and global COVID-19 specific patterns. The diverse boosted channels are achieved by employing STM blocks and transfer learning (TL) concepts to learn texture variation between COVID-19-specific and healthy images.

3. The COVID-19 infected CT images are provided to a novel Channel Boosted (CB) deep segmentation CNN, COVID-CB-RESeg, to precisely demarcate the COVID-19 infectious region in the lungs. To our knowledge, this is the first time we have incorporated CB in segmentation CNNs. The proposed COVID-CB-RESeg systematically employed region-homogeneity and heterogeneity operations in each encoder-decoder block and boosted-decoder in learning the low illumination and discriminative patterns of the COVID-19 infected region.

4. The performance comparison of novel SB-STM-BRNet detection and COVID-CB-RESeg segmentation CNN is on the standard SIRM COVID-19 CT dataset made with the existing CNNs.

The rest of the manuscript discusses the related work and proposed diagnosis framework in sections 2 and 3, respectively. Section 4 gives dataset, implementation, and performance metrics details. Section 5 provides insight into the result, discussion, and comparative analysis. Finally, section 6 outlines the manuscript conclusion and future work.

## 2. Related Work

Recently, CT technology has been used to diagnose COVID-19 infection in developed countries such as America, China, etc. However, manual examination of CT scans has a significant burden on radiologists and affects performance. Several researchers have focused on developing an automatic system to analyze suspected COVID-19 patients by analyzing CT scans [13]. Once confirmed that the suspected individual is COVID-19 infected, inform the close contacts, quarantine, and perform proper care and treatment of the infected person. Several classical techniques have been used for diagnosis but failed to show efficient performance [14]. Therefore, a deep CNN-based diagnostic automatic system has been developed for quick and accurate infection analysis to facilitate the radiologist [15]. This way, several deep CNNs, like VGG, ResNet, Xception, ShuffleNet, etc., have been employed on the COVID-19 dataset and achieved accuracy of 87% to 98% [16]–[18]. However, the aforementioned techniques have been employed for COVID-19 infection detection but lack analysis information.

Segmentation of contaminated regions, on the other hand, is commonly used to pinpoint the disease's location and severity. A deep CNN-based VB-Net has been designed to segment COVID-19 infection in CT scans and stated a 91.02 % dice similarity (DS). Joint-classification-segmentation framework has been employed to identify and analyze the infection and achieved 95% sensitivity, 93 % specificity, and a DS score (78.30%). Furthermore, the DCN method has been reported to analyze COVID-19 infection segmentation. The DCN method achieved an accuracy of 96.74% and a lesion segmentation DS of 83.51% [19]. The UNet and Feature Pyramid Network (FPN), with various encoder-backbone of DenseNet and ResNet, has been employed for lung region segmentation [20]. The reported technique showed a DS of 94.13%, an intersection over union (IoU) of 91.85%, and 99.64% sensitivity for the detection phase. Moreover, spatial and channel attention-based U-Net has been exploited spatially and channel-wise to capture rich contextual relationships to improve feature representation [21]. The obtained DS, detection rate, and Specificity are 83.1%, 86.7%, and 99.3%, respectively. A weakly-supervised DL technique is reported to perform lung segmentation and COVID-19 classification [22]. The reported technique achieved an accuracy of 96.2%, sensitivity of 94.5%, and DS = 90.0%. Moreover, Multi-task Learning based on encoder-decoder and MLP achieved an accuracy of 94.67%, and DS = 88.0% [23]. Furthermore, Infection Segmentation using Inf-Net CNN has been reported to segment the infectious regions in Lungs slices. The Inf-Net used a parallel partial decoder to enhance the high-level features and achieved a DS of 68.2% [24]. However, these diagnostic techniques have a few limitations:

1. Diagnosis is limited to detecting infected samples and, therefore, missing information about the disease's stages (mild, medium, and severe).

2. Most studies have employed existing deep CNN techniques, which may not be effective for COVID-19 analysis. These existing CNNs are explicitly developed for natural images, but the COVID-19 infectious region has different radiological patterns that are different from normal images. Therefore, there is a need to develop Deep CNN techniques to learn COVID-19 specific patterns, which may vary from other distorted or infected and healthy regions.

3. Patient data is usually imbalanced in nature and need to be evaluated using standard imbalanced measures such as F-score and MCC. Moreover, assessing the diagnostic technique using a few metrics provides unpredictable outcomes.

These challenges necessitate customizing the CNN architecture to exploit the region homogeneity, boundaries region, and textural variations associated with COVID-19-specific infection. Additionally, segmentation techniques are implemented to analyze the severity of the disease using stringent datasets and report the necessary performance metrics.

## 3. COVID-19 Diagnosis Framework

A new two-phase deep CNN-based system is proposed for automatically analyzing COVID-19 irregularities and specific lung radiographic patterns. COVID-19 specific radiographic patterns (region-homogeneity, textual variation, boundaries, etc.) are generally characterized by GGO, pleural effusion, consolidation, etc. Therefore, an integrated detection and segmentation framework is proposed that exploits COVID-19-specific radiographic patterns to identify and analyze COVID-19 infection in CT lung slices, as illustrated in Figure 1. The proposed framework has three main technical innovations, (i) the proposed SB-STM-BRNet detection model, (ii) the proposed COVID-CB-RESeg segmentation model, and (iii) the implementation of customized detection and segmentation CNNs. These customized CNNs have utilized training from scratch and TL for comparative analysis.

### 3.1. Proposed COVID-19 Infection Detection

A new deep CNN-based detection model, SB-STM-BRNet, is developed to discriminate COVID infectious images from healthy ones. The proposed detection phase constitutes two modules, the proposed SB-STM-BRNet, and customized existing CNNs. The COVID-19 detection setup is illustrated in Figure 2. Moreover, we have employed data augmentation in the detection and segmentation phase to improve the efficacy and reliability of real-time diagnostics.

### 3.1.1. Data Augmentation

COVID-19 is a new infectious disease, and publically available labeled CT datasets are limited. Therefore, the quantity of the dataset is increased by performing on-the-fly data augmentation at run-time during training models [25]. The technique is performed by applying the original image's rotation, shearing, reflection, etc., as depicted in Table 1.

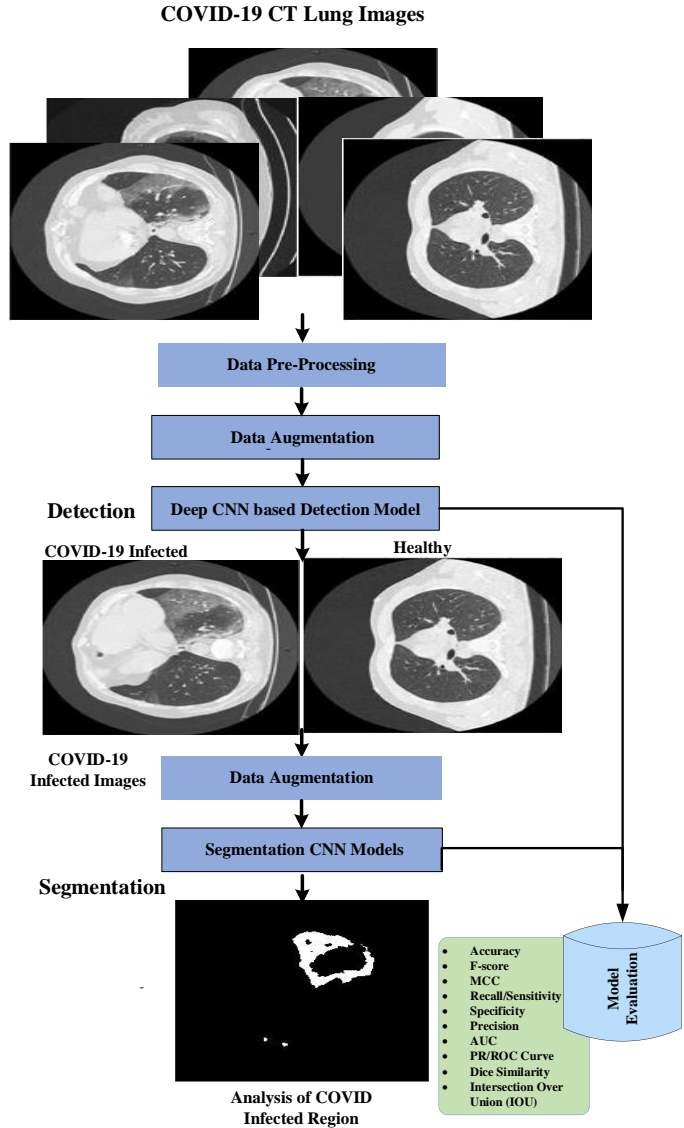

**Figure 1** The detailed workflow of the proposed COVID-19 diagnostic system. The workflow contains lung CT images, detection, segmentation models, and standard performance metrics.

### 3.1.2. The Proposed Channel SB-STM-RENet

The new SB-STM-BRNet deep CNN is developed, comprised of new dilated convolutional and STM blocks, and channel SB ideas to distinguish COVID-19 infected images from healthy slices. Each STM block systematically implements region and boundary operators within the dilated convolutional blocks. The SB-STM-BRNet comprises three dilated convolutional STM blocks with an identical topology where each block is composed of four regions, and boundary-based feature extraction dilated convolutional blocks. Each convolutional blocks methodically employed region and boundary operations using max and average-pooling appropriately [8], as shown in equations 2 and 3. These dilated convolutional blocks are named B, C, D, and E and are arranged methodically to learn various texture features and patterns in initial, intermediate, and final levels. In blocks D and E, additional channels are generated using TL to achieve various channels, while blocks B and C are original channels. The STM SB block dimensions are 32, 128, 256, and 128, 256, 512, respectively [26]. All four blocks are based on the STM technique to exploit the region, boundary, and global receptive feature extractions, the detailed architecture is shown in Figure 3.

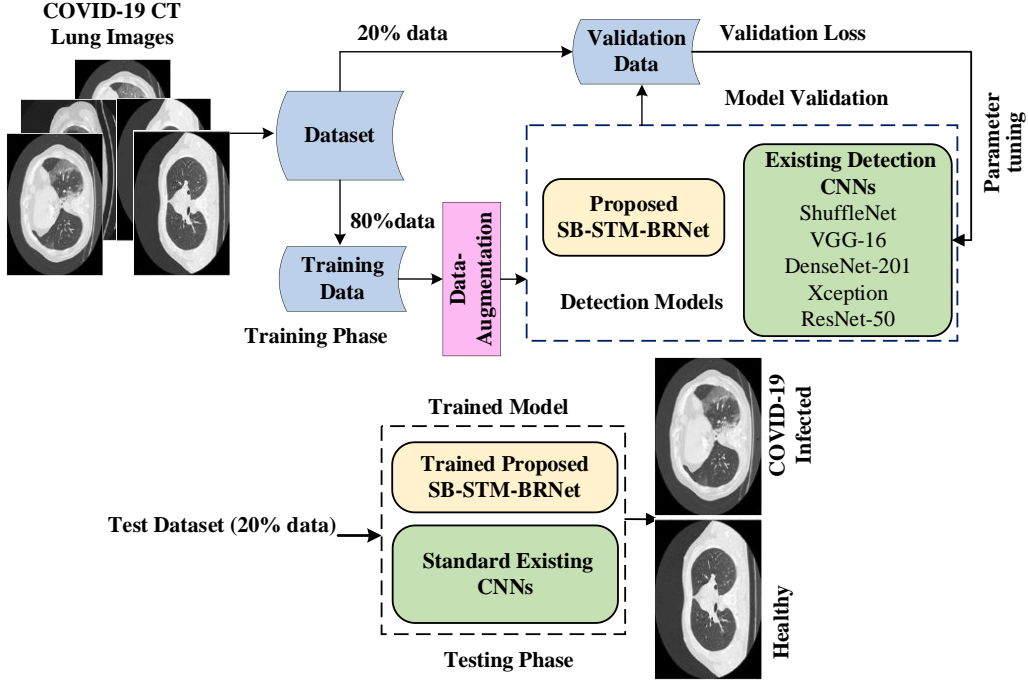

**Figure 2** The detailed workflow of the proposed COVID-19 Infection detection system. The workflow contains COVID-19 and healthy CT Lung images, detection models validation, and training

**Table 1** Data Augmentation techniques detail.

| Techniques | Values |
|------------|--------|
| Rotation | [± 30] angle |
| Shearing | [± 0.05] |
| Reflection | X, Y: [±1] |

$$x_{k,l} = \sum_{i=1}^{m} \sum_{j=1}^{n} x_{k+i-1,l+n-1} \, f_{i,j} \qquad (1)$$

$$x^{max}_{k,l} = max_{i=1,\ldots,w, j=1,\ldots,w} x_{k+i-1,l+j-1} \qquad (2)$$

$$x^{avg}_{k,l} = \frac{1}{w^2} \sum_{i=1}^{w} \sum_{j=1}^{w} x_{k+i-1,l+j-1} \qquad (3)$$

$$\mathbf{x}_{Boosted} = b(\mathbf{x}_B || \mathbf{x}_C || \mathbf{x}_D || \mathbf{x}_E) \qquad (4)$$

$$\mathbf{x} = \sum_{a}^{A} \sum_{b}^{B} y_a \, \mathbf{x}_{Boosted} \qquad (5)$$

$$\sigma(\mathbf{x}) = \frac{e^{x_i}}{\sum_{i=1}^{c} e^{x_c}} \qquad (6)$$

The channels and size are represented by **x** and k x l. The kernels and their size are denoted by **f** and i x j in equation 1. Equations 2 and 3 represent the average and max-pooling window size by **w**. In equations 4, the original feature map of block B and C is denoted by $\mathbf{x_B}$ and $\mathbf{x_C}$, respectively. Moreover, the auxiliary channel of block D and E achieved through TL is denoted as $\mathbf{x_D}$ and $\mathbf{x_E}$. $b(.)$ is the concatenation operations of original and auxiliary channels. The neuron quantity $\mathbf{y_a}$, softmax an activation function $\boldsymbol{\sigma}$ and $c$ represents classes, are presented in equations 5 and 6.

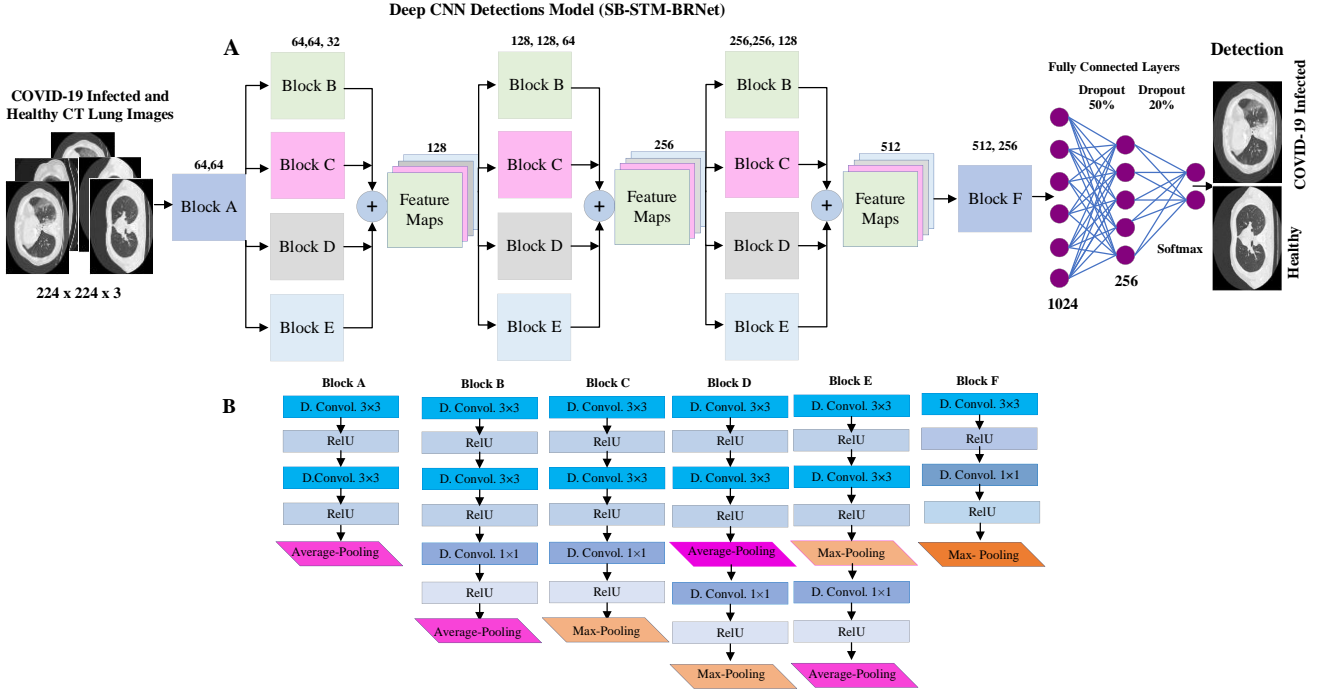

**Figure 3** The proposed SB-STM-BRNET CNN architecture for COVID-19 detection.

**Significance of New Channel SB-STM Blocks**

The proposed SB-STM block is based on the systematic use of dilated convolutional paired along with implementing average-pooling and max-pooling operations that enhance the feature set to distinguish healthy regions from the infected areas. Moreover, incorporating an edge operator helps the architecture learn highly discriminatory patterns, while the region operator for smoothening purposes. In addition, using various pooling operations results in down-sampling that eventually enhances the robustness of the model against any variation in input.

Furthermore, the perception of multipath-based STM blocks is used to obtain diversity in the feature set. The systematic stacking of STM blocks and pooling operations help to explore the COVID-19-specific patterns associated with region-homogeneity and boundary for various segments. Moreover, the new SB concept is incorporated at the STM blocks to get the prominent reduced channel and then combined to achieve various boosted diverse channels. Furthermore, the idea of the dilated convolutional layers is employed to preserve the global features. They can dynamically capture the representative information from the images and classify the CT Lung images using fully connected layers and dropout layers that preserve the prominent features and reduce overfitting.

## 3.2. Proposed COVID-19 Infection Analysis

Investigation of infected regions is essential to gain insights into the fundamental features of the infection pattern and its effects on the surrounding lung area. CT Lung images provide a higher level of detail in analyzing infected lung regions [16]. The CT lung images classified as COVID infected in the detection phase using the proposed SB-STM-BRNet are given to segmentation CNN as input. Moreover, data augmentation as preprocessing is employed to improve the model generalization.

### 3.2.1. The Proposed COVID-CB-RESeg Segmentation CNN

A novel COVID-CB-RESeg CNN is proposed to perform fine-grain pixel-wise segmentation to learn COVID-19-specific patterns. The proposed deep CNN-based COVID-CB-RESeg segmentation network comprises the novel boosted-decoder and the encoder blocks. The CB decoders are employed to preserve the spatial information of the corresponding encoders and auxiliary learners. The encoder and boosted-decoder blocks are designed in such a way to improve the COVID-CB-RESeg learning capacity. In this regard, average-pooling and max-pooling, along with convolutional operation in encoding stages, are employed systematically to learn region and boundary-related properties of COVID-19 infected regions [9]. Moreover, the convolutional operation employed a trained filter on images and generated feature maps of distinguishing patterns. The max-pooling is employed in the encoder for down-sampling to increase the robustness of the model and reduce the dimension. The up-pooling on the decoder side is initially employed up-sampling, and then novel channel boosting is incorporated. Finally, a 2x2 convolutional layer is employed to classify pixels into COVID-19 and background.

### The Systematic Exploitation of Regional and Edge Features

Max-pooling preserves boundary details for dissimilarity-based, while average-pooling stores regional information for pixel-wise object segmentation. In contrast, the systematic implementation of both average and max-pooling helps preserve hybrid and rich information feature sets to enhance the segmentation performance. These sets include region-homogeneity and boundary patterns of the COVID-19 infected region. Moreover, average-pooling suppresses noise acquired during CT image acquisition.

### Significance of Systematic Exploitation

The proposed strategy helps learn diverse feature sets and control the feature dimensions to improve generalization [8]. Normally, images are affected mainly due to illumination, contrast, and texture variation. Therefore, implementing both the pooling operations individually may affect the performance of CNNs. Therefore, max-pooling skips the region-homogeneity detail information while preserved by average-pooling. Moreover, average-pooling loss boundary information is recovered through max-pooling. The performance shows that the systematic utilization of region-homogeneity and boundary operations preserve rich information superior to individual processes. The main advantages of the proposed implementation are: (1) store region and smoothing details, (2) extracts high-intensity edge and boundary patterns as depicted in Equations 8 & 9. However, there may be a trade-off between information loss and pooling operations. Finally, $\mathbf{F}_{RE-e}$ and $\mathbf{F}_{RE-d}$ are region and boundary operations in encoders (e) and decoders (d) blocks of the proposed COVID-CB-RESeg CNN as shown in Equation 7 & 8, and Figure 5. Consequently, the CB operation at the decoder portion is illustrated in equation 9 and represented by $\mathbf{x}_{CB}$

$$\mathbf{F}_{RE-e} = f_c \left( \mathbf{x}^{avg}{}_{k,l} || \mathbf{x}^{max}{}_{k,l} \right) \tag{7}$$

$$\mathbf{F}_{RE-d} = f_c \left( \mathbf{x}^{max}{}_{k,l} || \mathbf{x}^{avg}{}_{k,l} \right) \tag{8}$$

$$\mathbf{x}_{CB} = b \left( \mathbf{F}_{RE-d} || \mathbf{x}_{AC} \right) \tag{9}$$

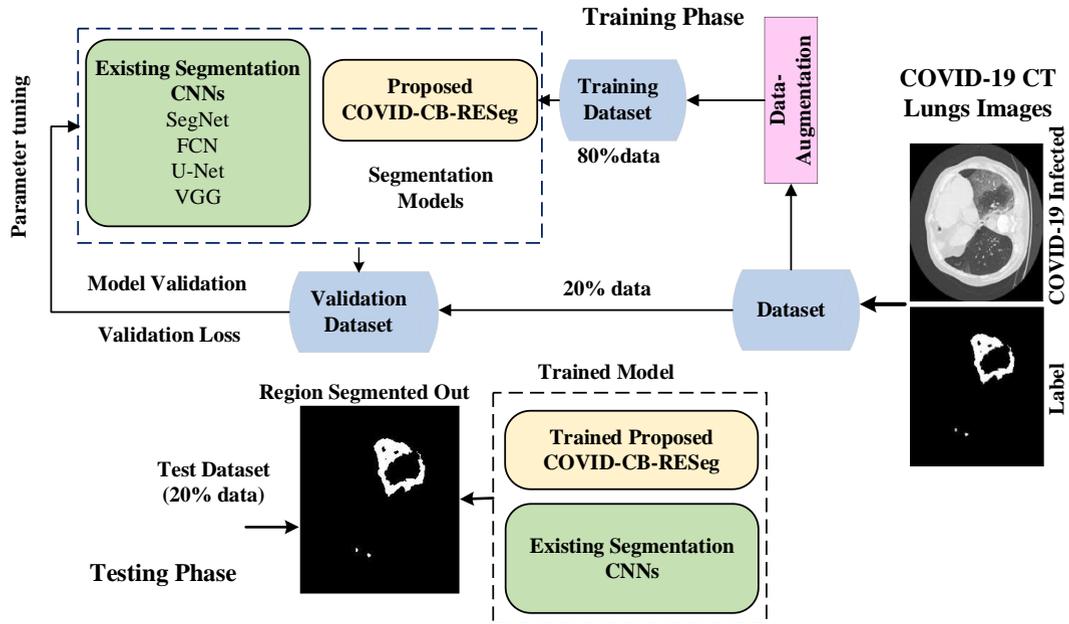

**Figure 4** The proposed COVID-RESeg CNN architecture for COVID-19 infection segmentation.

## Significance of Channel Boosting Utilization

In the proposed COVID-CB-RESeg segmentation CNN, the auxiliary decoder channels of TL-based customized CNNs are concatenated with the original channels of the proposed COVID-RESeg to attain diverse feature maps and improve the model's convergence. The COVID-RESeg are trained from scratch using the COVID-19 infected Lung dataset and exploiting CB using TL for persevering rich regional and edge-based information [27], [28]. The assistance of an auxiliary channel helps to enhance the proposed segmentation CNN representative's capacity. The proposed deep CNNs are predominantly based on the idea of TL to achieve significant performance. The weights are initialized from the TL-based pre-trained model's parameter space.

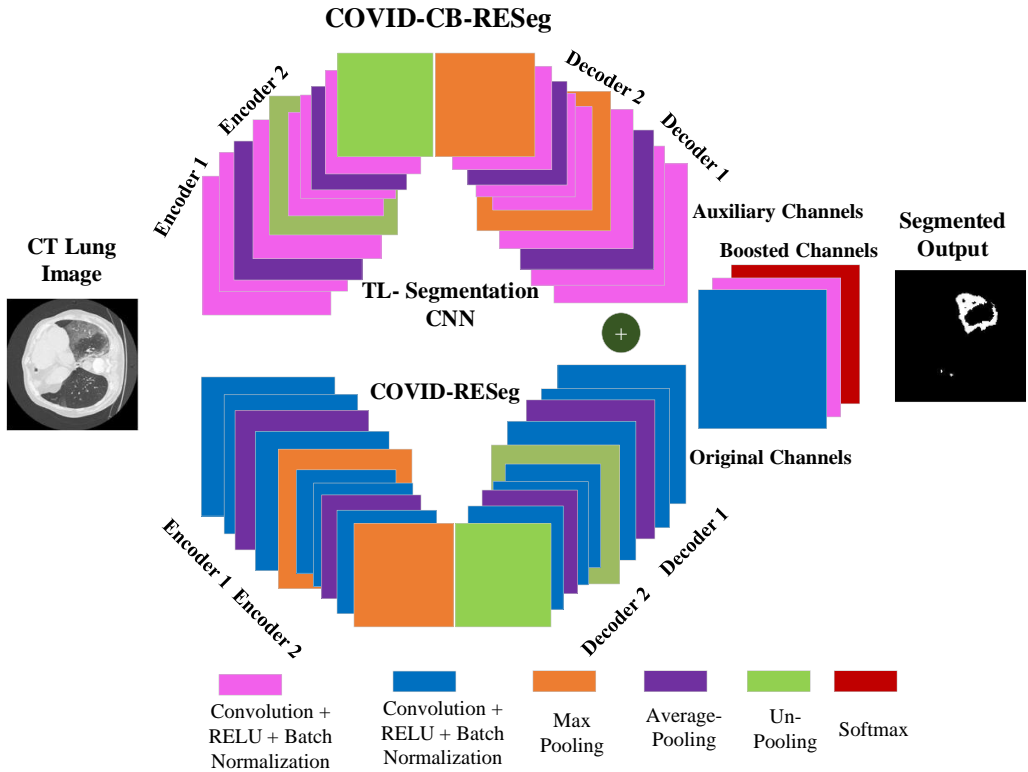

**Figure 5** The proposed COVID-CB-RESeg encoders, decoders, and the final layer of pixel-based segmentation.

### 3.3. Implementation of Existing Detection and Segmentation Models

Recently, CNN has demonstrated effective performance in medical field images to detect and segment medical images [29]. Various deep CNNs are employed to segment the COVID-19 CT infected region using diverse datasets [30], [31]. The existing deep detection and segmentation CNN has been implemented on the COVID-19 infected lung images for comparative studies. These models are customized by replacing the input and final pixel-label segmentation with new layers according to the data input size and output categories. We have employed the existing CNN models by training from scratch and weight initialization. The weights are initialized from pre-trained networks using the concept of TL and fine-tuned on CT images. The employed models for detection are VGG-16/19, ResNet-50, ShuffleNet, Xception, etc. [32]–[37] and VGG-16, SegNet, U-Net, and FCN as segmentation models [38], [39]. These segmentation models have different encoders, decoders, and other concepts of upsampling rate and skip connections.

## 4. Experimental Setup

### 4.1. Dataset

The lungs CT scan has a high sensitivity for the analysis of COVID-19 infection. The major benefit of using lung CT scans is that it makes the internal anatomy more apparent as overlapping structures are eliminated, thus leading to efficient analysis of the affected lung areas. The dataset of 30 patients with 2684 CT images is used provided by the Italian Society of Radiology (SIRM) [40]. The dataset is based on lung CT images and their corresponding labels on COVID-infected and healthy patients in .nii.gz format. The experienced radiologist examined the provided dataset, and the CT samples have paired with the binary label where infected lung regions are marked. Figure 6 illustrates the COVID-19 CT images, highlighting infected regions.

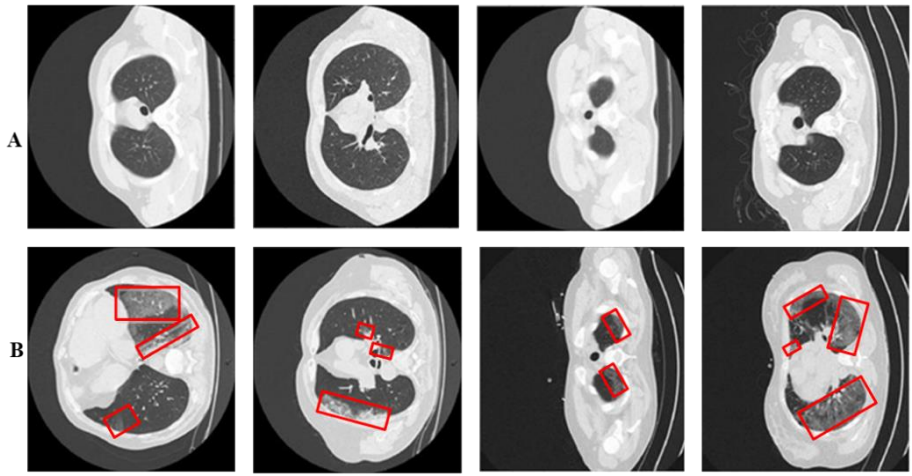

**Figure 6** Panel (A) and (B) illustrate Healthy vs. COVID-19 infected images. Moreover, the infectious regions are red-highlighted.

### 4.2. Data Pre-processing

The dataset is primarily based on 3D images in .nii.gz format; thus, it was converted into 2D images in .jpg format. This conversion divided each 3D CT lung image into several 2D slices. DL models tend to perform faster on small-sized images than on large-sized images. Thus, to boost the proposed diagnosis performance, the dataset images were resized from 512 x 512 x 3 to 304 x 304 x 3, which stemmed from the loss of information in segmentation. Resizing images is a critical aspect of preprocessing, which reduces the computation power and enhances training capacity [41].

### 4.2.1. Dataset Distributions for Detection and Segmentation phase

In the proposed framework, detection and segmentation networks are trained separately. The dataset consists of lung CT images of COVID infected and healthy patients. The dataset is categorized into two classes, COVID infected and healthy, depending upon their labels. The lung CT images of COVID and healthy classes are used in the proposed architecture for classification purposes. While for segmentation, only the COVID infected class images and their corresponding labels are used as it helps give a better insight into the infected area's analysis. The COVID-19 CT image datasets are divided into training (80%) and testing (20%) for both the detection and segmentation phases. The training set is further divided into training and validation sets using cross-validation techniques. The dataset distribution is available in Table 2.

**Table 2** Standard COVID-19 lung CT dataset description.

| Properties | Description |
|---|---|
| Total | 2684 Images |
| Original dimension | 224 x 224 x 3 |
| Healthy CT | 1362 Images |
| COVID-19 Infected CT | 1322 Images |
| Detection Training and Validation (80%) | (1717, 429) |
| Detection Testing (20%) | (538) |
| Segmentation Training and Validation (80%) | (846, 211) |
| Segmentation Testing (20%) | (265) |

## 4.3. Implementation Details

The hold-out cross-validation is used for optimal hyper-parameters selection and added validation data to the training portion. The optimal hyper-parameters used during training deep CNN for detection and segmentation is shown in Table 3. The proposed novel architecture based on DL detection and segmentation is built in the MATLAB 2021a tool. Since the training of the CNN network is computationally expensive, MatConvNet (a MATLAB-based DL library) was used for experiments [42]. The simulations are performed using NVIDIA GTX-T HP CPU and 32 GB RAM. Each model takes 1–2 hrs. ~10–20 min. /epoch, during training.

**Table 3** The hyper-parameters detail used during training deep CNN.

| Hyper-parameters | Values |
|---|---|
| Learning-rate | $10^{-3}$ |
| Epoch | 10 |
| Model-Optimizer | SGDM |
| Batch-size | 12 |
| Linear-Momentum | 0.90 |
| Loss-function | Cross-entropy |

## 4.4. Performance Evaluation

The proposed two-phase COVID-19 diagnosis system has been assessed using standard measures. Performance measures, abbreviations, and mathematical representation are detailed (Table 4). The detection measures are assessed using accuracy, recall, specificity, precision, MCC, and F-score and are mathematically expressed in Equations 10 to 15. Accuracy is the ratio of predictions that are correct out of all the predictions. Recall and precision are the correct ratio prediction of COVID-19 and Healthy slices, respectively. Finally, precision is the ratio of COVID-19 CT slices that are correctly predicted out of all the samples that are predicted as COVID-19 infected. Moreover, the segmentation measure includes IoU and the DS coefficient, as shown in Equations 16 & 17. The DS is the percentage of weighted similarity of the label and identified regions. In Equation 18, the standard error (S.E.) for DS score is computed at a 95% confidence interval (CI), where z=1.96 [43], [44]. The CI is used as a statistical test to evaluate the uncertainty of the segmentation CNNs.

**Table 4** Detail of detection and segmentation evaluation measures.

| Measure | Symbol | Explanation |
|---|---|---|
| **True-Positive** | TP | Correct predictions number for COVID-19 sample |
| **True-Negative** | TN | Total correct predictions of Healthy sample |
| **False-Positive** | FP | Total number of Healthy slices that are falsely COVID-19 detected |
| **False-Negative** | FN | Total COVID-19 slices that are falsely detected as Healthy |
| **Positive-Samples** | TP+FN | Entire COVID-19 slices |
| **Negative-Samples** | TN+FP | Entire Healthy slices |
| **True Prediction** | TP+TN | Total number of correctly classified slices |
| **Mathew Correlation Coefficient** | MCC | Identify the quality of confusion metrics on the unbalanced dataset. |
| **Negative Prediction** | FP+FN | Total number of misclassified samples |
| **Jaccard Coefficient** | IoU | %Similarity between the original label and predicted region |
| **Dice Similarity** | DS | %Weighted similarity between the original label and predicted region |
| **Segmentation Acc** | S-Acc | %Pixels that are correctly segregated to the positive Infected and Background. |

$$\text{Accuracy} = \frac{\text{Detected COVID19} + \text{Detected healthy}}{\text{Total samples}} \quad (10)$$

$$\text{Precision} = \frac{\text{Detected COVID19}}{\text{Detected COVID19} + \text{Incorrectly detected COVID19}} \quad (11)$$

$$\text{Recall} = \frac{\text{Detected COVID19}}{\text{Total COVID19 samples}} \quad (12)$$

$$\text{Specificity} = \frac{\text{Correctly detected healthy}}{\text{Total healthy}} \quad (13)$$

$$\text{F} - \text{Score} = 2\frac{(\text{P x R})}{\text{P} + \text{R}} \quad (14)$$

$$\text{MCC} = \frac{(\text{TP x TN}) - (\text{FN x FP})}{\sqrt{(\text{TP} + \text{FP})(\text{FP} + \text{FN})(\text{TN} + \text{FP})(\text{TN} + \text{FN})}} \quad (15)$$

$$\text{IoU} = \frac{\text{Truely segmented infectious region}}{\text{Truely segmented infectious region} + \text{Entire infectious region}} \quad (16)$$

$$\text{DS} = \frac{2 * \text{Truely segmented infectious region}}{2 * \text{Truely segmented infectious region} + \text{Entire infectious region}} \quad (17)$$

$$\text{CI} = z\sqrt{\frac{error(1 - error)}{Total\ instances}} \quad (18)$$

## 5.   Results and Analysis

This paper proposes a new two-phase diagnosis framework to detect and analyze the COVID-19 infectious region in the lungs. The two-phase framework has the advantages of improving diagnosis performance and reducing computational complexities. Moreover, this process rivals the clinical workflow, where patients are referred for further diagnosis after initial detection, which helps quickly identify the severity of the disease. The proposed framework is tested on unseen data and indicates exceptional performance compared to existing architectures. Moreover, a data augmentation strategy has been applied to the trained dataset, showed beneficial results, and enhanced the model generalization.

### 5.1.   Performance of COVID-19 Detection

The detection phase is optimized to recognize the COVID-19 infectious patterns and reduced false negatives. Therefore, a novel SB-STM-BRNet CNN is developed to detect COVID-19 infectious images in the initial phase. Moreover, the learning capacity of the SB-STM-BRNet and customized CNNs for COVID-19 features is evaluated on test data.

### 5.1.1.   Performance Analysis of the Proposed SB-STM-BRNet

The proposed SB-STM-BRNet achieves reasonable generalization compared to existing good-performing ResNet-50 in terms of F-score (SB-STM-BRNet: 98.24 %, ResNet-50: 94.93 %), accuracy (SB-STM-BRNet: 98.21 %, ResNet-50: 94.35 %), and MCC: (SB-STM-BRNet: 94.85 %, ResNet-50: 89.89%), as illustrated in the Table 5. The learning plot for the proposed detection SB-STM-BRNet CNN is shown in Figure 7 and demonstrates training and validation accuracy. The proposed boundary and region-based STM blocks enhance the detection rate by classifying maximum samples as true positives. SB-STM-BRNet learned homogeneity and boundary-based COVID-19 specific patterns in the STM block. Moreover, the proposed SB-STM-BRNet incorporated the idea of channel SB and TL that improved the sensitivity and maintained a high F-score in Table 5 and Figure 8. Furthermore, the new SB achieved prominent and diverse feature maps learned small illumination changes and texture variation among COVID-19 infected and healthy regions in CT images [45]. However, fewer miss-classified images are due to contaminants or noise in CT images, which yield a likeness between COVID-19 and healthy people [46].

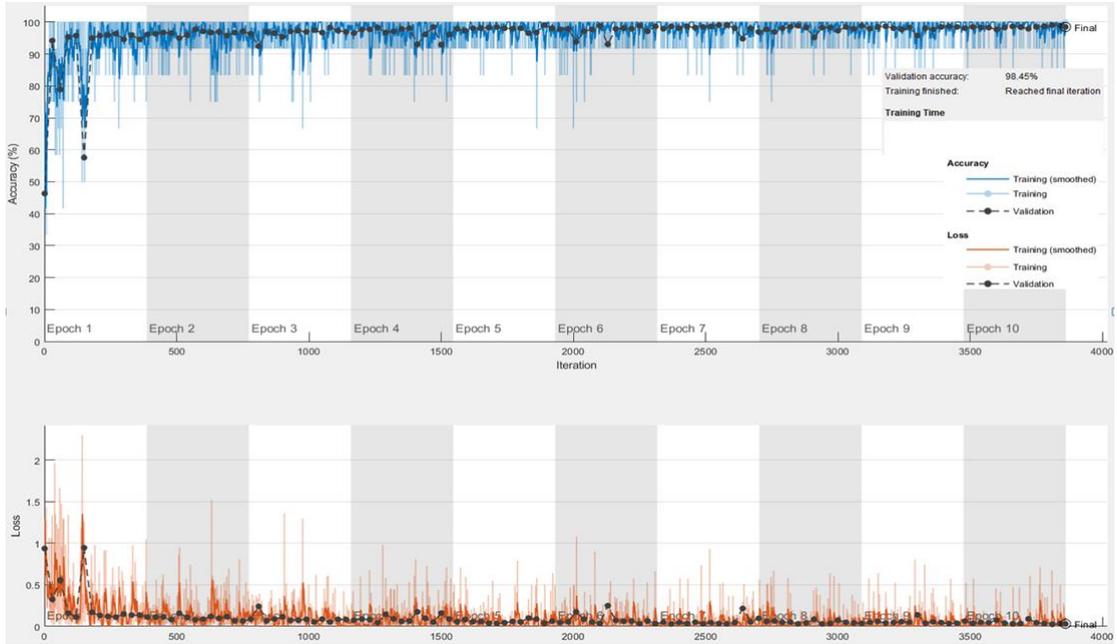

**Figure 7** Convergence plot during training of the proposed SB-STM-BRNet detection CNN.

### 5.1.2. Performance Comparison with the Existing Detection CNNs

The proposed SB-STM-BRNet performance is compared with the five customized classification CNNs (ShuffleNet, VGG-16, DenseNet-201, Xception, and ResNet-50). Customized CNNs are famous for solving complex challenges and maybe successively used to identify lung abnormalities. SB-STM-BRNet shows outperformance with the customized existing CNNs on the test dataset, and a fair comparison is shown in Table 5. TL-based channel generation, channel squeezing, and boosting to achieve prominent and diverse feature maps are considered efficient strategies in deep CNN architecture to enhance COVID-19 detection performance. The proposed SB-STM-BRNet achieved the performance gain over the existing CNNs regarding Accuracy = (1.48 - 8.33 %), F-score = (1.47 – 8.24%), MCC = (2.83 – 14.89%), and Recall = (1.18-7.06%), as shown in Table 5 and Figure 8. Moreover, TL-based existing CNN improved performance as compared to trained from scratch regarding Accuracy = (2.38 %), F-score = (2.44%), MCC = (1.91%), and Recall = (2.35%). Thus, TL enhances SB-STM-BRNet model generalization and convergence because of pre-trained optimized weights and then fine-tuned on COVID-19 specific patterns [47]. TL has demonstrated impressive performance in medical diagnosis challenges, especially COVID-19 and cancer diagnosis.

**Table 5** Performance comparison of the proposed SB-STM-BRNet and existing techniques.

| Model | Accuracy | F-score | ROC-AUC | PR-AUC | Precision | MCC | Specificity | Recall |
|-------|----------|---------|---------|--------|-----------|-----|-------------|--------|
| ShuffleNet | 89.88 | 90.00 | 96.40 | 96.36 | 88.85 | 79.76 | 88.55 | 91.18 |
| TL-ShuffleNet | 92.26 | 92.44 | 96.40 | 96.36 | 91.38 | 81.87 | 90.96 | 93.53 |
| VGG-16 | 92.86 | 92.86 | 98.33 | 98.84 | 92.78 | 85.71 | 92.77 | 92.94 |
| TL-VGG-16 | 94.35 | 94.43 | 98.33 | 98.84 | 94.15 | 87.21 | 93.98 | 94.71 |
| DenseNet-201 | 94.35 | 94.23 | 99.09 | 98.58 | 96.82 | 88.87 | 96.99 | 91.67 |
| TL-DenseNet-201 | 95.83 | 95.81 | 99.09 | 98.58 | 97.56 | 90.67 | 96.99 | 94.12 |
| Xception | 94.64 | 94.54 | 98.45 | 98.93 | 96.84 | 89.44 | 96.99 | 92.35 |
| TL-Xception | 96.13 | 96.12 | 98.45 | 98.93 | 97.58 | 91.52 | 97.59 | 94.71 |
| ResNet-50 | 94.94 | 94.93 | 99.26 | 98.76 | 95.16 | 89.89 | 95.18 | 94.71 |
| TL-ResNet-50 | 96.73 | 96.77 | 99.26 | 98.76 | 96.49 | 92.02 | 96.39 | 97.06 |
| **Proposed SB-STM-BRNet** | **98.21** | **98.24** | **99.67** | **99.10** | **98.24** | **94.85** | **98.19** | **98.24** |
| JCS [48] | --- | --- | --- | --- | --- | --- | 93.00 | 95.00 |
| VB-Net [16] | --- | --- | --- | --- | --- | --- | 90.00 | 87.00 |
| DCN [19] | --- | 96.74 | --- | --- | --- | --- | --- | --- |
| 3DAHNet[49] | | | | | | | 90.00 | 85.00 |

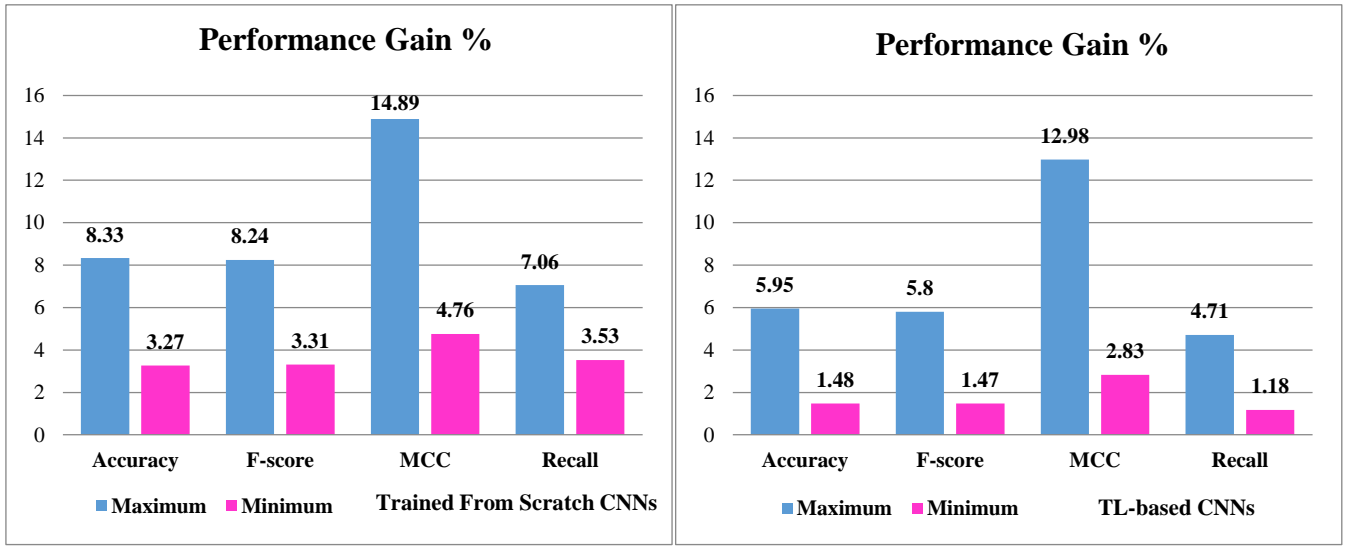

**Figure 8** The proposed SB-STM-BRNet performance gain over the existing detection CNN regarding standard performance metrics.

### 5.1.3. Feature Space Analysis

The feature space learned by the proposed SB-STM-BRNet is visualized to recognize the decision-making behavior more efficiently. Usually, the distinguishable class features improve the learning capacity and reduce the model variance using a stringent dataset. The 2-D scatter plot of the principal components (PC) and their percentage variance on test data for the existing good-performing ResNet-50 are shown in Figure 9. Moreover, the 2-D scatter plot of the PC1 Vs. PC2 and PC1 Vs. PC3 and their percentage variance for SB-STM-BRNet are shown in Figure 10. Finally, the feature space of the proposed SB-STM-BRNet visualizes good discrimination between COVID-19 and healthy features, which is illustrated in Figure 10.

### 5.1.4. Detection Significance of the Proposed SB-STM-BRNet

Detection rate curves are used to quantitatively assess the discrimination ability of the developed SB-STM-BRNet. These are performance measurement curves that evaluate the generalization of the SB-STM-BRNet by analyzing the discrimination between two COVID-19 infected and healthy classes at different threshold setups. Moreover, Figure 11 shows evidence from ROC and PR curves that SB-STM-BRNet has a good learning ability compared to existing detection CNNs on unseen data.

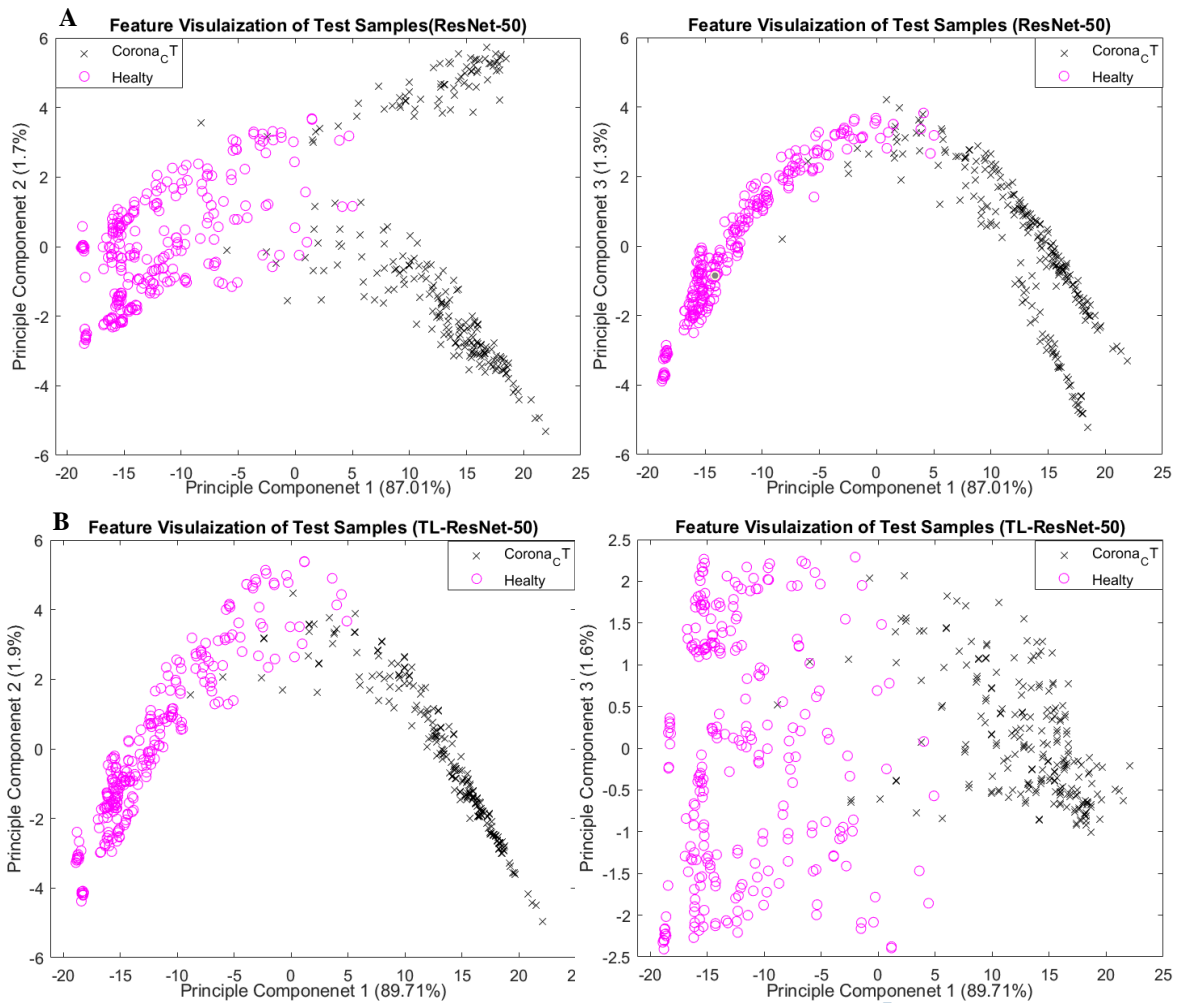

**Figure 9** Visualization of first, second, and third PCs of feature-space generated; Panel (A) illustrates the best performing state-of-the-art ResNet-50, while Panel (B) shows TL-based ResNet-50, respectively).

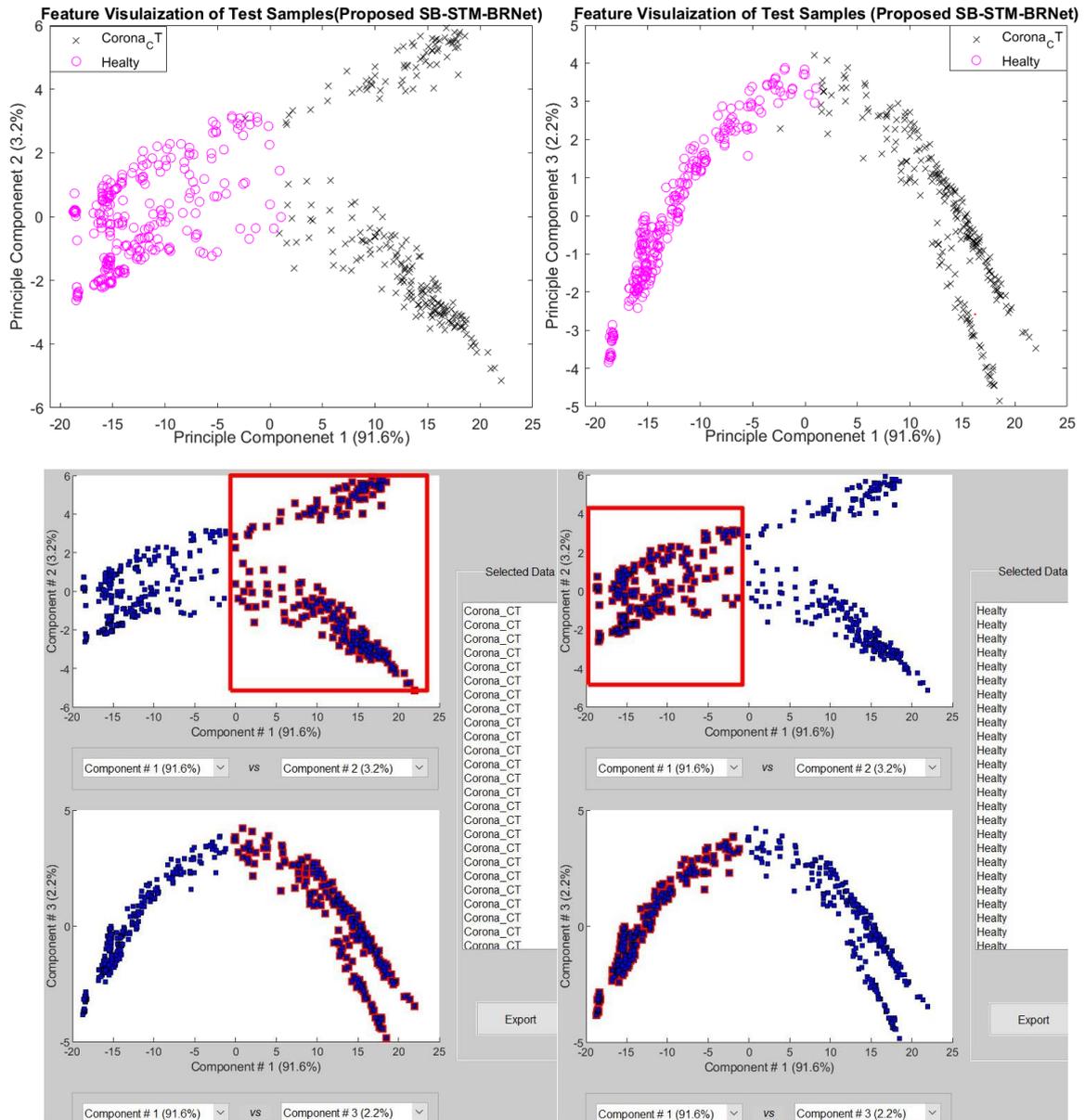

**Figure 10** Feature Space Visualization of the proposed SB-STM-BRNet for the generated first, second, and third PCs.

## 5.2. COVID-19 Infection Analysis

CT infected images are screened using the proposed SB-STM-BRNet provided to deep segmentation CNN for infectious region analysis. The analysis of the infected region is required for an insight view of infectious patterns and their implication on surrounding or other organs. Moreover, region analysis is needed to identify the severity of the mild, medium, or severe disease and its treatment design. The COVID-CB-RESeg is developed to segment COVID-19 infectious regions in CT lung images.

In the proposed COVID-CB-RESeg, the new concept of region-homogeneity and boundary-based implementation is methodically incorporated using average-pooling and max-pooling. The systematic implementation helps learn well-defined boundaries and segregates the infected regions by identifying infection boundaries with minor contrast variation. Moreover, TL-based channel generation and boosting improved the proposed CNN learning ability to capture texture variation of the infectious region at the target level. The results suggest that the proposed COVID-CB-RESeg has an excellent learning ability for COVID-19 infectious patterns, evident from DS (96.40 %) and IoU (98.85 %), respectively (Table 6). Consequently, precisely learned the discriminative boundaries and achieved a higher value of BFS (99.09 %). The proposed COVID-CB-RESeg appears globally suited for moderate to severely infected regions.

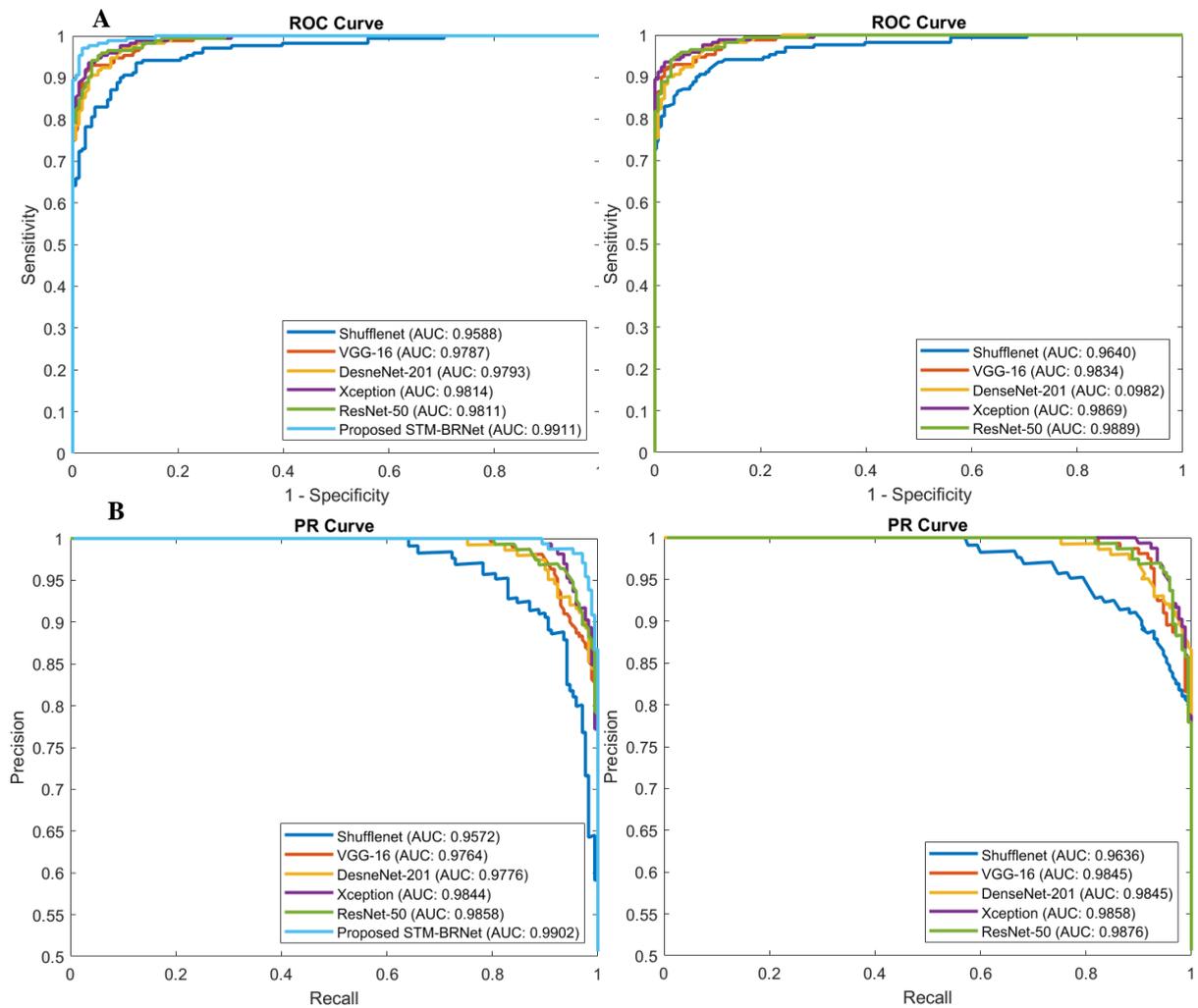

**Figure 11** ROC and PR curve-based analysis of the proposed SB-STM-BRNet and well-established customized detection CNNs.

**Table 6** The proposed COVID-CB-RESeg comparison with existing techniques and segmentation CNNs on unseen data. The S.E. for DS indices is calculated at 95% CI.

| CNN | Region | DS ± S.E. | Acc. | IOU | BFS | G_Acc. | M_Acc. | M_IOU | W_IOU | M_BFS |
|---|---|---|---|---|---|---|---|---|---|---|
| COVID-CB-RESeg | COVID-19 | **96.40±4.70** | **99.21** | **98.85** | **99.09** | **99.51** | **99.49** | **98.98** | **99.09** | **98.32** |
| | Background | **99.70±1.40** | **99.72** | **99.31** | **97.45** | | | | | |
| TL-SegNet | COVID-19 | 93.50±6.90 | 98.36 | 97.61 | 96.88 | 99.11 | 99.01 | 98.04 | 98.14 | 96.01 |
| | Background | 96.80±4.90 | 99.86 | 98.47 | 96.82 | | | | | |
| SegNet | COVID-19 | 93.20±6.90 | 98.01 | 97.21 | 96.5 | 96.01 | 98.81 | 98.66 | 97.64 | 96.01 |
| | Background | 96.60±4.90 | 99.51 | 98.07 | 96.44 | | | | | |
| TL-UNet | COVID-19 | 94.10±5.90 | 98.29 | 97.60 | 97.41 | 99.10 | 98.95 | 98.13 | 98.22 | 96.82 |
| | Background | 97.10±3.50 | 99.62 | 98.45 | 96.23 | | | | | |
| U-Net | COVID-19 | 93.60±5.40 | 97.94 | 97.2 | 97.03 | 96.82 | 98.8 | 98.6 | 97.73 | 96.82 |
| | Background | 96.70±3.10 | 99.27 | 98.05 | 95.85 | | | | | |
| TL-VGG-16 | COVID-19 | 93.00±7.50 | 98.61 | 95.88 | 96.91 | 98.40 | 98.44 | 96.70 | 96.87 | 95.49 |
| | Background | 96.70±4.30 | 98.28 | 97.32 | 94.07 | | | | | |
| VGG-16 | COVID-19 | 92.80±7.80 | 98.26 | 95.48 | 96.53 | 95.49 | 98.1 | 98.09 | 96.3 | 95.49 |
| | Background | 96.30±4.80 | 97.93 | 96.92 | 93.69 | | | | | |
| TL-FCN-8 | COVID-19 | 89.70±11.30 | 90.92 | 89.01 | 87.74 | 95.32 | 94.55 | 90.63 | 90.20 | 82.43 |
| | Background | 93.00±7.50 | 98.18 | 92.05 | 77.11 | | | | | |
| FCN-8 | COVID-19 | 89.20±11.90 | 90.57 | 88.61 | 87.36 | 82.43 | 95.02 | 94.2 | 90.23 | 82.43 |
| | Background | 92.50±7.90 | 97.83 | 91.65 | 76.73 | | | | | |
| FPN[20] | Infected | 0.941±0.059 | --- | 91.85 | --- | --- | --- | --- | --- | --- |
| VB-Net [16] | COVID-19 | 0.910±0.090 | --- | --- | --- | --- | --- | --- | --- | --- |
| Weakly Sup.[22] | COVID-19 | 90.00+10.5 | --- | --- | --- | --- | --- | --- | --- | --- |
| Multi-stask Learning [23] | COVID-19 | 88.0+12.7 | --- | --- | --- | --- | --- | --- | --- | --- |
| DCN [19] | Infected | 0.835±0.175 | --- | --- | --- | --- | --- | --- | --- | --- |
| U-Net-CA [21] | COVID-19 | 83.10±16.9 | --- | --- | --- | --- | --- | --- | --- | --- |
| Inf-Net [24] | COVID-19 | 0.682±0.329 | --- | --- | --- | --- | --- | --- | --- | --- |

### 5.2.1. Comparative Analysis of Segmentation CNNs

The proposed COVID-CB-RESeg performance is compared with four popular segmentation CNNs (SegNet, U-Net, VGG-16, and FCN). The learning ability and the segmented infected regions by COVID-CB-RESeg and existing CNNs are shown in Table 6, Figure 12, and Figure 13. The proposed COVID-CB-RESeg achieved performance gain over the existing CNNs regarding DS (2.8- 7.2%), IoU (1.64 - 10.24 %), and BFS score (2.06 - 11.73 %), as depicted in Table 6. Furthermore, the pixel-wise segmentation of the infection region and created maps using COVID-CB-RESeg illustrate high subjective quality compared to existing CNNs (Figure 13). The performance metrics and the visual quality of the segmented maps are evidence that the proposed COVID-CB-RESeg accurately highlights the infected region. In existing CNNs, SegNet offers good performance with a DS score of 93.50 %, IoU of 97.21 %, and BFS of 96.50 %. However, FCN fluctuates in learning infectious patterns at various stages of COVID-19, fails to retain infected regions, and thus shows less robustness.

### 5.2.2. TL based Segmentation Analysis

The TL-based performance gain over the trained from scratch is approximately 0.40 to 1%, as shown in Table 6 and Figure 12. TL enhances the model convergence and generalization by achieving optimized weights or learned patterns from pre-trained scenarios. Incorporating pixel-wise distribution of the proposed COVID-CB-RESeg improved the segmentation for different stages of infections. The proposed COVID-CB-RESeg has low complexity and in-depth but shows more accurate performance than high complex and large depth models.

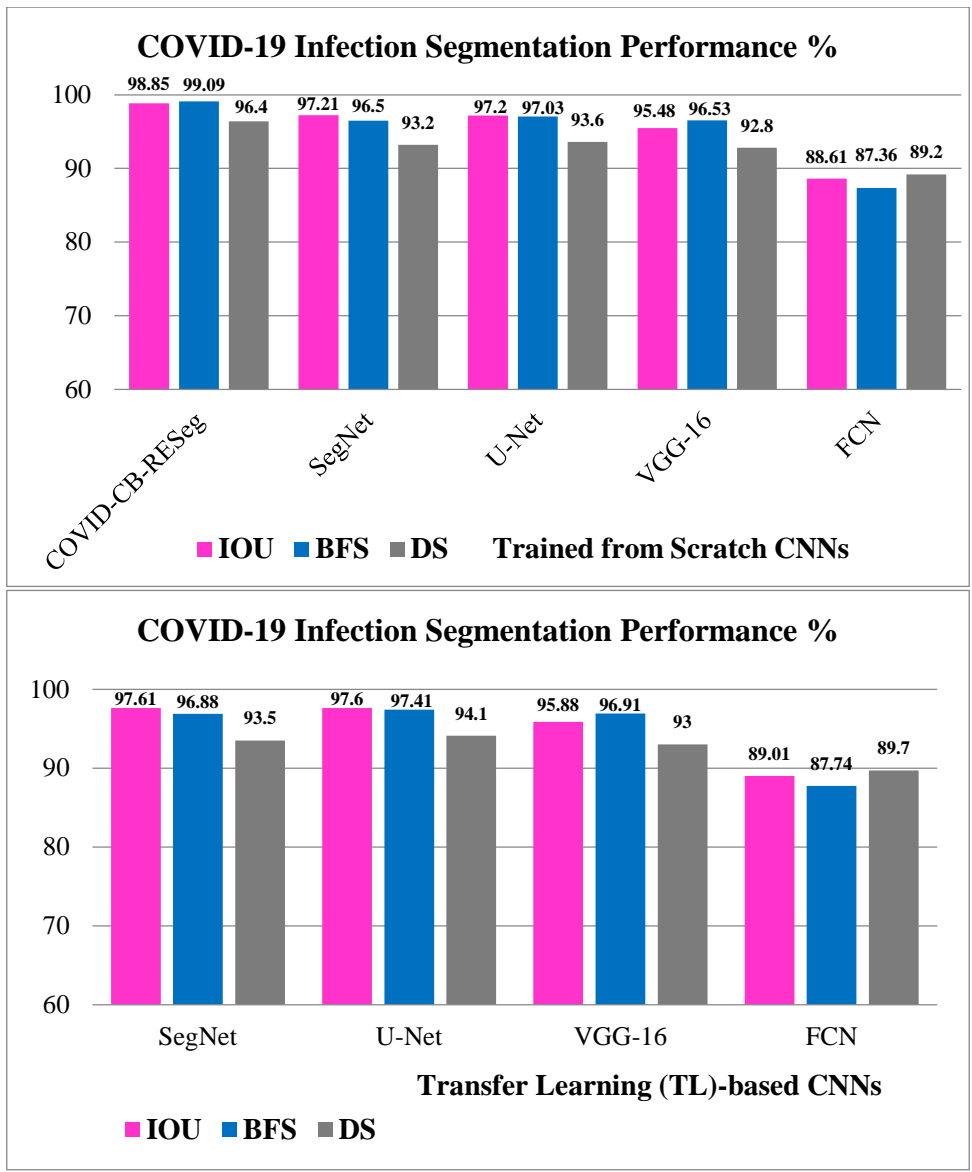

**Figure 12** Quantitative evaluation of the proposed COVID-CB-RESeg, TL-based fine-tuned and trained from scratch CNNs for COVID-19 infection segmentation.

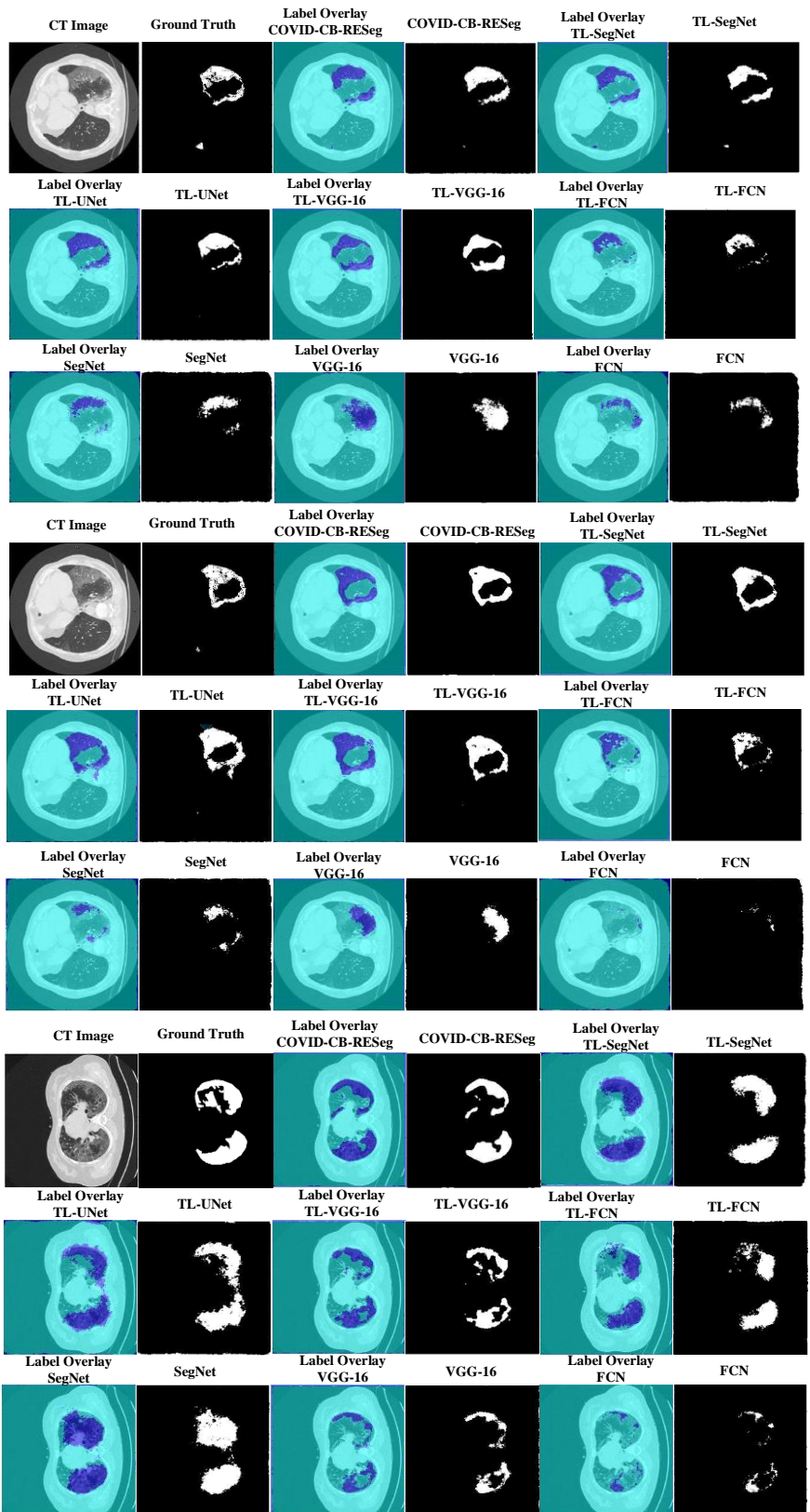

**Figure 13** Visual evaluation of the proposed COVID-CB-RESeg, fine-tuned using TL and training from scratch segmentation CNNs. The visual results illustrate the original CT image, label, label overlay, and segmented results of segmentation CNN.

## 6. Conclusion

COVID-19 is a transmissible disease that has primarily affected worldwide. Therefore, an early diagnosis of COVID-19 infection is required to control the spread of the disease. A new two-phase deep CNN-based framework is proposed to screen COVID-19 infection and analyze COVID-19 contagious regions in CT lungs to identify the severity of the disease (mild, medium, and severe). The novel SB-STM-BRNet detection and COVID-CB-RESeg CNN benefited from data augmentation and boosting using TL-based rich information channels generation and concatenation. Moreover, SB-STM-BRNet benefited from channel squeezing and new dilated convolutional STM blocks for achieving prominent feature maps and diverse global feature sets. The proposed SB-STM-BRNet achieved the highest performance, with an accuracy of 98.21 % for segregating COVID-19 instances from healthy individuals with an F-score: of 98.24 % and an MCC: of 94.85 %. Moreover, the outperformance of the proposed COVID-CB-RESeg (IoU: 98.85%, DS: 96.40%) on the test dataset shows that it can precisely analyze the infected regions in CT lung images. The proposed framework exploits homogenous regions, textural variations, and boundaries in images which help differentiate the COVID-19 infected local and global regions from healthy. The proposed diagnostic system's considerable performance suggests that an integrated approach is appropriate for the early screening of COVID-19 infection and detailed analysis of the COVID-19 infectious region. Moreover, the quick and computer-aided diagnosis will help save valuable lives and have good socio-economic impacts. In the future, the proposed diagnosis framework will be extended to the multi-class challenge (COVID-19, viral and bacterial pneumonia, and Healthy) and other types of lung abnormalities. GAN will be used to augment the dataset by generating synthetic examples to improve efficacy and reliability for real-time diagnostics.

**Acknowledgment**

We thank the Department of Computer Systems Engineering, University of Engineering and Applied Sciences (UEAS), Swat, for providing the necessary computational resources and a healthy research environment.